\documentclass[aps,twocolumn,showpacs,superscriptaddress,prb]{revtex4}

\usepackage[dvips]{graphicx}
\usepackage[fleqn]{amsmath}
\usepackage{amsfonts}
\usepackage{amssymb} 
\usepackage{subfigure}
\usepackage{epsfig}
\usepackage{color}

\newcommand{\Turbo}{\textsc{Turbomole}}
\newcommand{\omat}[1]{\boldsymbol{#1}}

\begin{document}

\title{\emph{Ab initio} study of charge transport through 
single oxygen molecules in atomic aluminum contacts}

\author{S. Wohlthat}
\altaffiliation{Present address: 
School of Chemistry, The University of Sydney, Sydney,
New South Wales 2006, Australia}
\affiliation{Institut f\"{u}r Theoretische Festk\"{o}rperphysik and 
DFG-Center for Functional Nanostructures,
Universit\"{a}t Karlsruhe, D-76128 Karlsruhe, Germany}
\affiliation{Institut f\"{u}r Nanotechnologie, 
Forschungszentrum Karlsruhe, D-76021 Karlsruhe, Germany}
\author{F. Pauly}
\affiliation{Institut f\"{u}r Theoretische Festk\"{o}rperphysik and 
DFG-Center for Functional Nanostructures,
Universit\"{a}t Karlsruhe, D-76128 Karlsruhe, Germany}
\affiliation{Institut f\"{u}r Nanotechnologie, 
Forschungszentrum Karlsruhe, D-76021 Karlsruhe, Germany}
\author{J. K. Viljas}
\affiliation{Institut f\"{u}r Theoretische Festk\"{o}rperphysik and 
DFG-Center for Functional Nanostructures,
Universit\"{a}t Karlsruhe, D-76128 Karlsruhe, Germany}
\affiliation{Institut f\"{u}r Nanotechnologie, 
Forschungszentrum Karlsruhe, D-76021 Karlsruhe, Germany}
\author{J. C. Cuevas}
\affiliation{Institut f\"{u}r Theoretische Festk\"{o}rperphysik and 
DFG-Center for Functional Nanostructures,
Universit\"{a}t Karlsruhe, D-76128 Karlsruhe, Germany}
\affiliation{Institut f\"{u}r Nanotechnologie, 
Forschungszentrum Karlsruhe, D-76021 Karlsruhe, Germany}
\affiliation{Departamento de F\'{i}sica Te\'{o}rica 
de la Materia Condensada C-V, Universidad Aut\'{o}noma de Madrid, 
E-28049 Madrid, Spain}
\author{Gerd Sch\"{o}n}
\affiliation{Institut f\"{u}r Theoretische Festk\"{o}rperphysik and 
DFG-Center for Functional Nanostructures,
Universit\"{a}t Karlsruhe, D-76128 Karlsruhe, Germany}
\affiliation{Institut f\"{u}r Nanotechnologie, 
Forschungszentrum Karlsruhe, D-76021 Karlsruhe, Germany}
\date{\today}

\begin{abstract} 
  We present \emph{ab initio} calculations of transport properties of
  atomic-sized aluminum contacts in the presence of oxygen. 
  The experimental situation is modeled by considering a 
  single oxygen atom (O) or one of the molecules O$_2$ and O$_3$
  bridging the gap between electrodes forming ideal, 
  atomically sharp pyramids.  
  The transport characteristics are computed for these
  geometries with increasing distances between the leads, simulating
  the opening of a break junction. To facilitate comparison with
  experiments further, the vibrational modes of the oxygen connected to
  the electrodes are studied. It is found that in the contact
  regime the change of transport properties due to the presence of
  oxygen is strong and should be detectable in experiments. All
  three types of oxygen exhibit a comparable behavior in their
  vibrational frequencies and conductances, which are well below 
  the conductance of pure aluminum atomic contacts. The conductance
  decreases for an increasing number of oxygen atoms. In the
  tunneling regime the conductance decays exponentially with
  distance and the decay length depends on whether or not oxygen is
  present in the junction. This fact may provide a way to identify the
  presence of a gas molecule in metallic atomic contacts.
\end{abstract}

\maketitle

\section{Introduction\label{chap:intro}}
Since the first experiments on atomic contacts, the ``atmospheric''
surroundings are known to have a strong effect on their charge
transport properties.\cite{Gimzewski1987} Thus, studying the influence
of adsorbed gas molecules is of great importance for the design of
functional molecular-scale electronic components. It also shows some
promise for the future development of gas sensors for ``electronic
noses''. However, the systematic experimental investigation of the
influence of single gas molecules on charge transport through metallic
atomic contacts started only a few years ago with hydrogen in atomic
platinum contacts.\cite{Smit2002,Djukic2005} It was found that the
zero-bias conductance through such platinum-hydrogen contacts tends to
be close to the quantum of conductance $G_0=2e^2/h$.  In addition, 
this conductance value is due to a single, almost fully transparent 
channel.\cite{Smit2002,Djukic2006}
Theoretical work has mostly confirmed these
findings.\cite{Djukic2005,Cuevas2003,Garcia2004,Thygesen2005,Garcia-Suarez2005}

Even so, one of the main problems in experiments of this type is to
know if and in which configuration gas molecules are present in the
contact region.  A definite identification of the presence and the
type of gas can be based on statistical techniques, such as the
measurement of conductance histograms.\cite{Smit2002,Agrait2003} For
individual contacts, this becomes much harder. In the experiments of
Refs.\ \onlinecite{Smit2002,Djukic2005} the identification of a single
hydrogen molecule was based on the so-called inelastic point contact
spectroscopy, where the local vibrational modes can be seen in the
current-voltage ($I$-$V$) characteristics as abrupt features at bias
voltages corresponding to the energies of the
vibrations.\cite{Agrait2003} The evolution of these features with
stretching of the contacts can then be compared to calculations of the
vibrational modes for given configurations of the
molecule.\cite{Smit2002,Djukic2005} In fact, theoretical methods have
recently been developed that enable a direct comparison between the
full experimental and theoretical $I$-$V$ characteristics, including
the inelastic signatures of the vibrational modes.
\cite{Frederiksen2004,Viljas2005,delaVega2006,Ness2006,Solomon2006,Frederiksen2006}

For this purpose, namely to identify small molecules in atomic metal
contacts through transport measurements, we study the general
properties of oxygen (O) in atomic aluminum (Al) contacts.  In this
paper we concentrate on the behavior of the zero-bias transport characteristics
when the atomic contacts are stretched. In the spirit of Refs.\
\onlinecite{Smit2002,Garcia2004,Djukic2005}, we also study the
evolution and character of the vibrational modes.  Aluminum was
chosen, because it becomes superconducting below temperatures of 1 K.
As a result, in addition to measurements of the total zero-bias
conductance, it is possible to determine experimentally the
transmission eigenvalues of the individual channels from the $I$-$V$
characteristics in the superconducting state.\cite{Scheer1997} As
particular examples, we study single oxygen atoms (O), oxygen dimers
(O$_{2}$), and ozone (O$_{3}$).  These molecules (we also refer to the
oxygen atom as a molecule) are placed initially in the middle of the
contact between two aluminum electrodes, since their effect on the
transport properties is then at its largest. We describe our contacts at the 
level of density-functional theory (DFT). In particular, the DFT electronic 
structure is employed to determine the charge transport properties by 
means of Green's function techniques.

In the following we present stable structures of the three types of
contacts, where oxygen molecules bridge the gap between two idealized aluminum
electrodes. For them, we calculate the vibrational modes and the
low-temperature transport properties, including the characterization
of the individual transmission channels.  We find that both the
energies of the vibrational modes, as well as the evolution of the
conductance with increasing electrode distance show qualitatively
comparable characteristics for the three investigated molecules.
In contrast to pure Al, the conductances of the Al-O$_x$-Al
contacts are always well below $1 G_0$, and for an increasing number of O
atoms, the conductance decreases. Moreover, the exponential decay
lengths in the tunneling regime are different in the presence and in
the absence of oxygen.

This paper is organized as follows. In Sec.\ \ref{chap:method} we begin
by presenting the methods used for determining the electronic structure,
the geometries, and the transport properties of the molecular
contacts. In Sec.\ \ref{chap:modes} we then describe the optimized
geometries and the corresponding vibrational modes. Following this,
Sec.\ \ref{chap:transport} relates in detail the transmission
properties of the junctions to the relevant molecular orbitals. The
section ends with an analysis of the evolution of the conductances when 
the contacts are opened. In particular, we analyze the tunneling regime,
which is discussed further in App.\ \ref{app.a}. In Sec.\ \ref{chap:discussion}
we conclude with the discussion of our results.

\begin{figure*}
\begin{center}
\begin{minipage}{0.99\linewidth}
\subfigure[Al-O-Al\label{fig:Ostruct}]{
\includegraphics[width=0.3\linewidth]{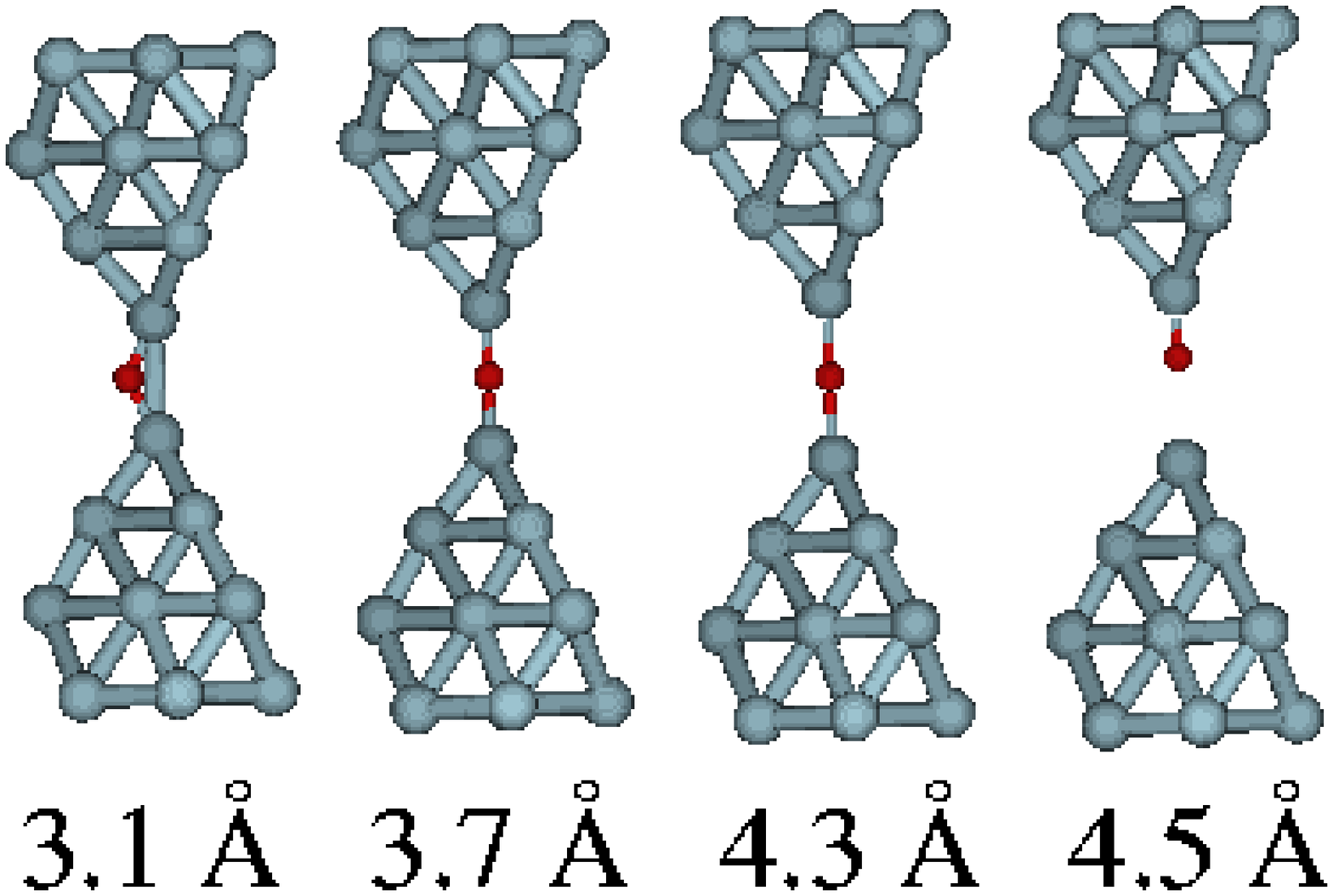}
}
\vspace{0.01cm}
\subfigure[Al-O$_{2}$-Al\label{fig:O2struct}]{
\includegraphics[width=0.3\linewidth]{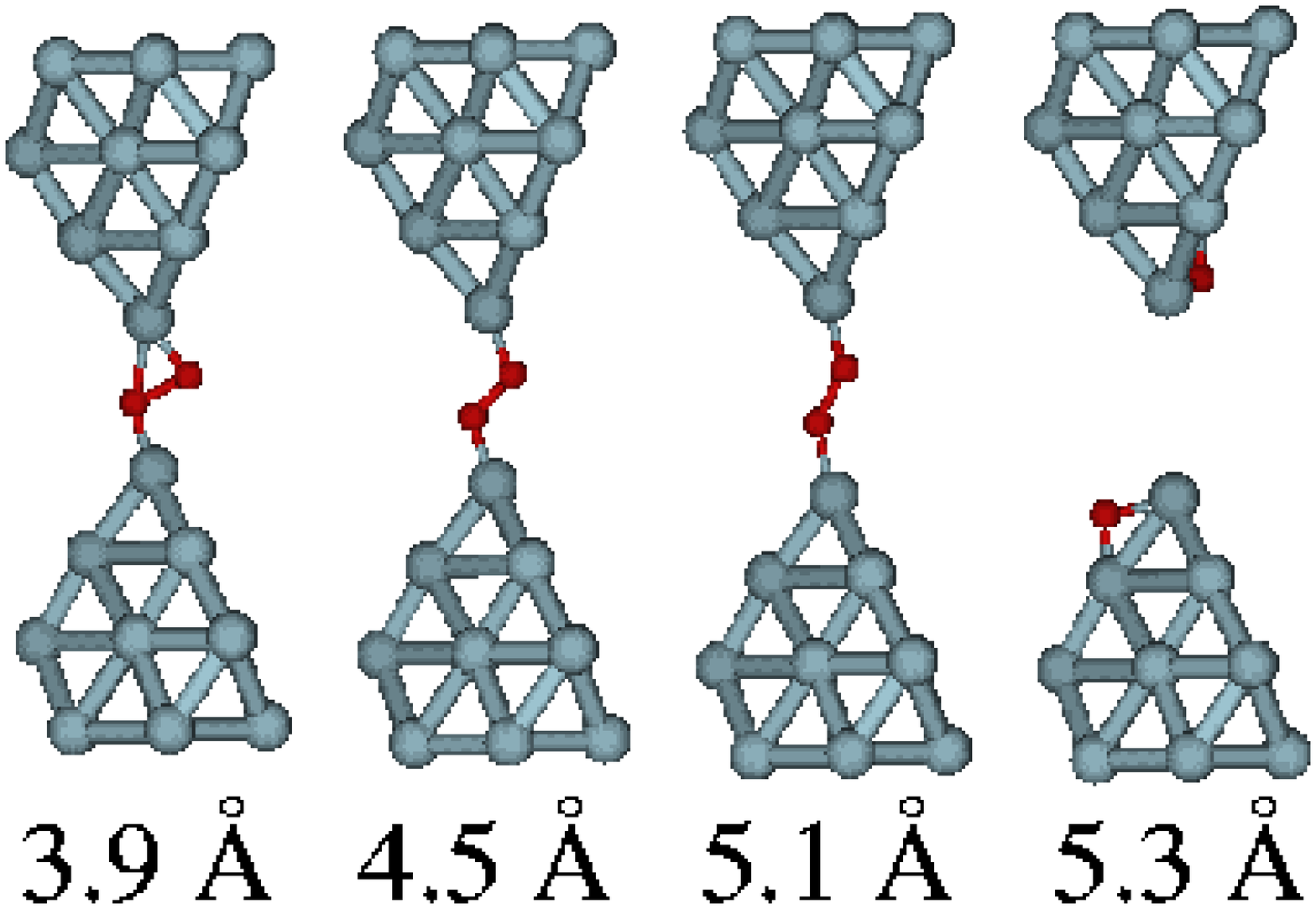}
}
\vspace{0.01cm}
\subfigure[Al-O$_{3}$-Al\label{fig:O3struct}]{
\includegraphics[width=0.3\linewidth]{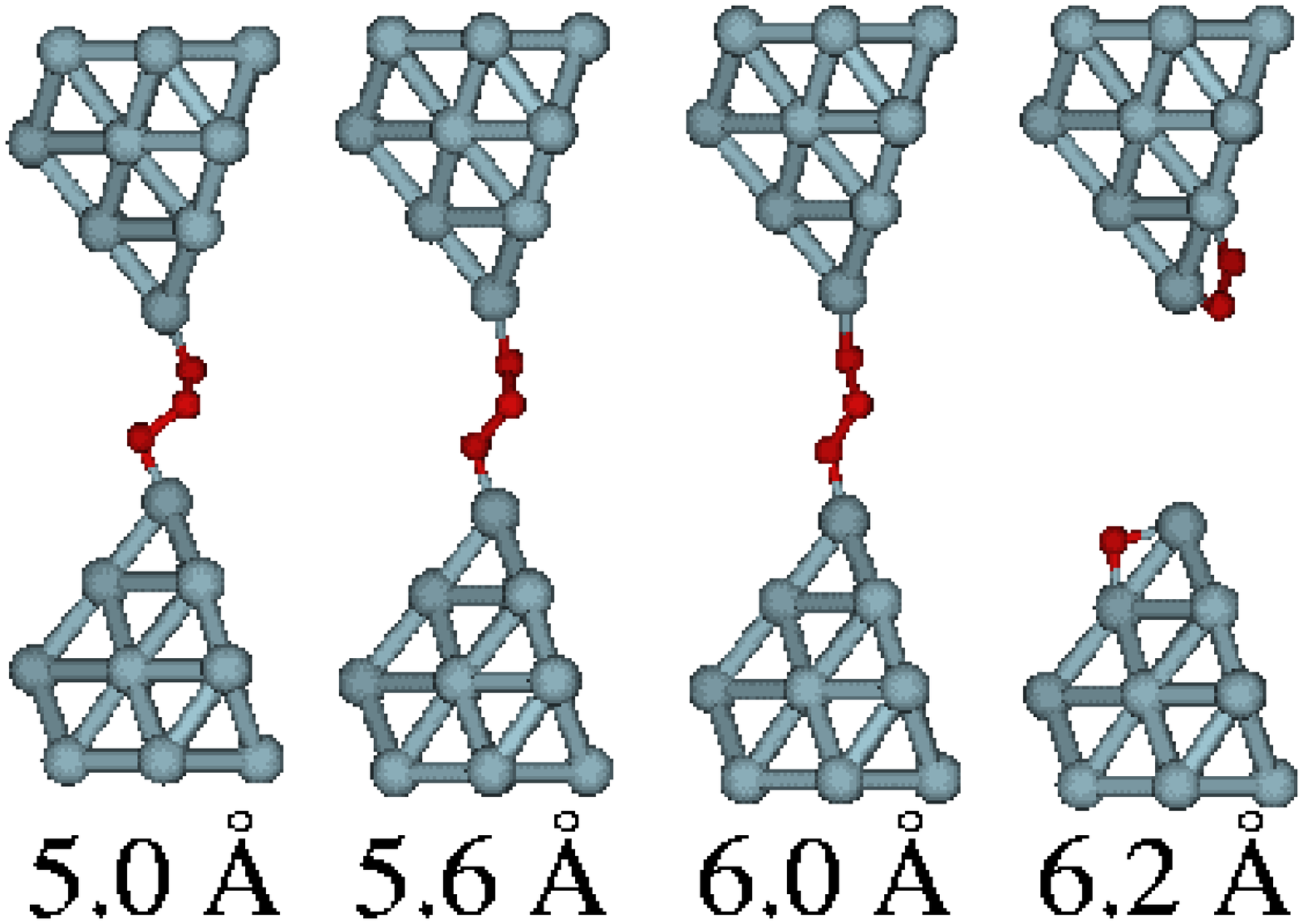}
}
\caption[Al$_{17}$-O$_{x}$-Al$_{17}$: Chosen relaxed 
structures from the simulation of the opening of an aluminum (blue) 
point contact in 
the presence of oxygen (red) molecules.]
{\label{fig:Struktur}
(Color online)
Al$_{17}$-O$_{x}$-Al$_{17}$: Optimized contact geometries and their 
tip-to-tip distances as obtained by stretching an aluminum (blue)
atomic contact in the presence of oxygen (red) molecules.
 The Al pyramids consist of 7, 6, 3, and 1 
atoms in the layers from the outside to the inside.}
\end{minipage}
\end{center}
\end{figure*}

 \begin{table*}
 \begin{center}
 \begin{minipage}{0.98\linewidth}
 \subtable[Al-O-Al]{
 \begin{tabular}{|*{3}{c|}} \hline
 Al-Al & Al-O & O-Al \\\hline
 3.097 \AA & 1.700 \AA & 1.700 \AA \\
 3.297 \AA & 1.685 \AA & 1.685 \AA \\
 3.497 \AA & 1.749 \AA & 1.749 \AA \\
 3.697 \AA & 1.849 \AA & 1.849 \AA \\
 3.897 \AA & 1.948 \AA & 1.948 \AA \\
 4.097 \AA & 2.048 \AA & 2.048 \AA \\
 4.297 \AA & 2.150 \AA & 2.147 \AA \\
 4.497 \AA & 2.779 \AA & 1.718 \AA \\\hline
 \end{tabular}
 }
 \subtable[Al-O$_{2}$-Al]{
 \begin{tabular}{|*{4}{c|}} \hline
 Al-Al & Al-O & O-O & O-Al \\\hline
 3.938 \AA & 1.812 \AA & 1.571 \AA & 1.796 \AA \\
 4.138 \AA & 1.785 \AA & 1.535 \AA & 1.785 \AA \\
 4.338 \AA & 1.783 \AA & 1.503 \AA & 1.787 \AA \\
 4.538 \AA & 1.799 \AA & 1.486 \AA & 1.803 \AA \\
 4.738 \AA & 1.828 \AA & 1.488 \AA & 1.830 \AA \\
 4.938 \AA & 1.867 \AA & 1.511 \AA & 1.869 \AA \\
 5.138 \AA & 1.900 \AA & 1.571 \AA & 1.899 \AA \\\hline
 \end{tabular}
 }
 \subtable[Al-O$_{3}$-Al]{
 \begin{tabular}{|*{5}{c|}} \hline
 Al-Al & Al-O & O-O & O-O & O-Al \\\hline
 4.970 \AA & 1.798 \AA & 1.471 \AA & 1.461 \AA & 1.802 \AA \\
 5.170 \AA & 1.802 \AA & 1.464 \AA & 1.456 \AA & 1.806 \AA \\
 5.370 \AA & 1.809 \AA & 1.461 \AA & 1.454 \AA & 1.813 \AA \\
 5.570 \AA & 1.820 \AA & 1.463 \AA & 1.456 \AA & 1.822 \AA \\
 5.770 \AA & 1.833 \AA & 1.469 \AA & 1.464 \AA & 1.835 \AA \\
 5.970 \AA & 1.855 \AA & 1.486 \AA & 1.476 \AA & 1.858 \AA \\\hline
 \end{tabular}
 }
 \label{tab:distances} 
 \caption{ Distances between the aluminum tip atoms and the lengths of
   the individual Al-O, O-O, and O-Al bonds for the contact geometries
   depicted in Fig.\ \ref{fig:Struktur}. The order from top to bottom
   corresponds to an order from left to right in the figure.  }
 \end{minipage}
 \end{center}
 \end{table*} 

\section{Method\label{chap:method}}

In this section we briefly present our methods for computing the
electronic structures, geometries, vibrational modes, as well as the 
charge transport characteristics of atomic contacts. A more detailed
account will be given elsewhere.\cite{Method2006} 

We describe our contacts at the level of density functional theory
(DFT) as implemented in the quantum chemistry package \Turbo\ v5.7
(Refs.\
\onlinecite{Ahlrichs1989,Treutler1994,Hohenberg1964,Kohn1965}).  In
particular, we use the ``ridft''
module\cite{Eichkorn1995,Eichkorn1997} with the BP86 exchange
correlation functional\cite{Becke1988,Perdew1986} and a Gaussian basis
set of split valence plus polarization (SVP)
quality.\cite{Schafer1992} To be precise, polarization functions are
present on all non-hydrogen atoms within the employed basis set. The
total energy is converged to a precision of $10^{-6}$ Hartree.
Geometry optimization is performed until the maximum norm of the
Cartesian gradient has fallen below $10^{-4}$ atomic units.
Vibrational modes in our contacts are determined using the ``aoforce''
module of \Turbo.\cite{Deglmann2002-1,Deglmann2002-2,Horn1991}

The investigated atomic contacts consist of two atomically sharp
electrode tips and a single oxygen molecule connecting them
(Fig.~\ref{fig:Struktur}).  The electrodes are Al fcc pyramids
oriented in the (111) direction with a lattice constant of 4.05~\AA\
and consist of at least four layers. While the outer layers are kept
fixed all the time, the inner two layers of the freestanding pyramids
are relaxed to find an energy minimum. Afterwards the atoms in the Al
pyramids are kept fixed.  Between two such electrodes we place one
representative oxygen molecule (O, O$_{2}$, or O$_{3}$).  The molecule
is then relaxed between the fixed electrodes, leading to a stable
configuration. Starting from this first configuration, we move the
electrodes stepwise further away from each other and relax the oxygen
molecule again in each step. For each relaxed structure we determine
the vibrational modes of the molecule connected to the leads.

To compute the charge transport through a molecule between metallic
electrodes, we apply a method based on standard Green's function
techniques and the Landauer formula expressed in terms of a local 
non-orthogonal basis.
\cite{Taylor2001,Damle2001,Brandbyge2002,Xue2002,Xue2003,Viljas2005} 
Due to the locality of the basis,
the system is conveniently separated into three parts, leading to a
single-particle Hamiltonian (or Fock matrix) of the molecular junction
that has the following form
\begin{equation}
\omat{H}=
\left(\begin{matrix}
\omat{H}_{LL} &\omat{H}_{LC} &\omat{H}_{LR} \\
\omat{H}_{CL} &\omat{H}_{CC} &\omat{H}_{CR} \\
\omat{H}_{RL} &\omat{H}_{RC} &\omat{H}_{RR} 
\end{matrix}
\right).
\end{equation}
Here $\omat{H}_{CC}$ describes the central system ($C$) consisting of
the molecule and the tips of the electrodes and $\omat{H}_{XX}$ with
$X=L,R$ describe the left ($L$) and right ($R$) electrodes.  The
matrices $\omat{H}_{LC}=\omat{H}_{CL}^T$ etc.\ give the hopping
elements between the different subsystems, but the hoppings between
$L$ and $R$ are assumed to vanish
($\omat{H}_{LR}=\omat{H}_{RL}=\omat{0}$).  This assumption has been
verified in all our calculations, where the Hamiltonian elements
connecting $L$ and $R$ were always smaller than $10^{-6}$ Hartree.
The overlap matrix $\omat{S}$ of the non-orthogonal basis has a
similar structure.  

The conductances of the molecular junctions are determined by means
of Green's function techniques. The low-temperature zero-bias
conductance $G=dI/dV|_{V=0}$ is given by the Landauer formula
\begin{equation}\label{e.landauer}
G=G_0\textrm{Tr}[\omat{t}(E_F)\omat{t}^\dagger(E_F)]=G_0T(E_F),
\end{equation}
where $\omat{t}(E_F)$ is the transmission matrix of the molecular
junction evaluated at the Fermi energy $E_F$.  The total transmission
at any energy $E$ can be written as $T(E)=\sum_{n=1}^{\infty}T_n(E)$, where
$T_n(E)$ are the transmissions of the individual eigenchannels,
defined as the eigenvalues of $\omat{t}(E)\omat{t}^\dagger(E)$.  The
transmission matrix can be calculated in terms of Green's functions of
the molecular junction as follows\cite{Heurich2002}
\begin{equation}
\omat{t}(E)=2\omat{\Gamma}^{1/2}_L(E)\omat{G}^r_{CC}(E)\omat{\Gamma}^{1/2}_R(E).
\end{equation}
Here
\begin{equation}
\omat{G}^r_{CC}(E)=[E\omat{S}_{CC}-\omat{H}_{CC}
-\omat{\Sigma}^r_L(E)-\omat{\Sigma}^r_R(E)]^{-1}
\end{equation}
is the retarded Green's function of the central system, and
$\omat{\Gamma}_{X}(E)=-\textrm{Im}[\omat{\Sigma}^r_{X}(E)]$
are the scattering rate matrices.
These depend on the self-energy matrices
\begin{equation} \label{e.selfenergy}
\omat{\Sigma}_{X}^r(E)=(\omat{H}_{CX}-E\omat{S}_{CX})
\omat{g}_{XX}^r(E)(\omat{H}_{XC}-E\omat{S}_{XC}),
\end{equation}
where $\omat{g}_{XX}^r(E)$ is the retarded surface Green's
function of the electrode $X=L,R$, which are modeled as 
ideal surfaces. These self-energies contain all 
the information about the electronic structure of the electrodes 
and their coupling to the central system.

The electronic structure of the central system and its coupling to 
the electrodes are obtained from the calculated geometries (Fig.~\ref{fig:Struktur}).
Let us refer to these geometries, which consist of the two pyramids bridged by 
a molecule, as ``contact geometries''. These contact geometries are 
divided into three parts. The central part, corresponding to
region $C$ in the transport calculation, consists of the molecule and
the inner two layers of the pyramids. The excluded outer layers on the
left and right sides are used for calculating the hoppings and
overlaps between the central region and the electrodes as needed in
Eq.\ (\ref{e.selfenergy}).  

To obtain the surface Green's functions $\omat{g}_{XX}^r(E)$, we
compute separately the electronic structure of a spherical Al fcc
cluster with 555 atoms.  From this we extract the Fock and overlap
matrix elements between the atom in the origin of the cluster and all
its neighbors and, using these ``bulk parameters'', construct a
semi-infinite periodic crystal.  The surface Green's functions are
then calculated from this crystal using the so-called decimation
technique.\cite{Guinea1983} We have checked that the electrode
construction in the employed non-orthogonal SVP basis set has
converged with respect to the size of the Al cluster, from which we
extract our parameters.\cite{Method2006} In this way we describe the
whole system consistently within DFT, using the same basis set and
exchange-correlation functional everywhere.

The Fermi energy $E_F$ of the coupled system, namely the $C$ part
connected to the semi-infinite electrodes, required to determine the
conductance [Eq.\ (\ref{e.landauer})], is assumed to be given by the
Al leads.  We obtain its value as the average of the energies of the
highest occupied ($-4.261$ eV) and the lowest unoccupied ($-4.245$ eV)
orbitals of the bulk cluster.  In this way we find $E_F=-4.25$ eV.
The same quantity for the contact geometries deviates from this bulk
value by less than $0.08$ eV. Let us stress that the Fermi energy is
only used to read off the conductance from a transmission function
$T(E)$, but does not enter into our calculations otherwise.

\section{Contact geometries and vibrational modes \label{chap:modes}}

\begin{figure*}
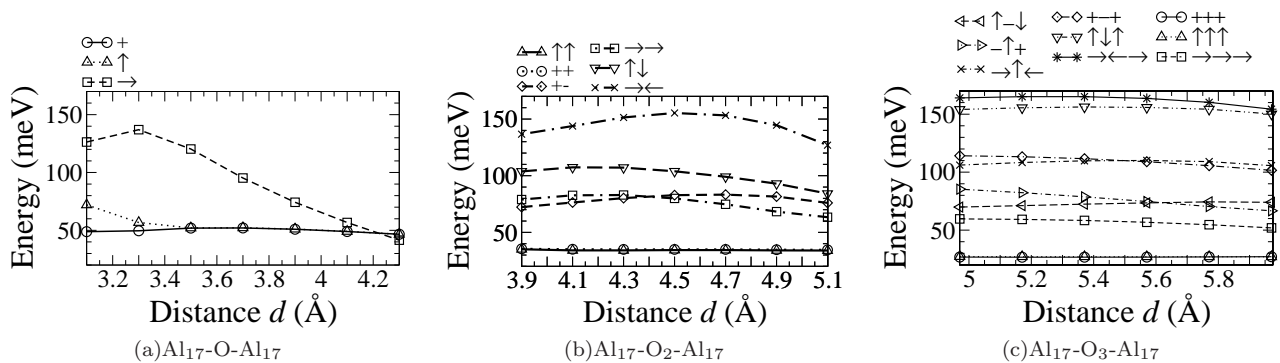

\begin{center}
\begin{minipage}{0.98\linewidth}
\subfigure[Al$_{17}$-O-Al$_{17}$\label{fig:Ovib}]{
\includegraphics[width=0.3\linewidth]{VibrationsO}
}
\hspace{0.1cm}
\subfigure[Al$_{17}$-O$_{2}$-Al$_{17}$\label{fig:O2vib}]{
\includegraphics[width=0.3\linewidth]{VibrationsO2}
}
\hspace{0.1cm}
\subfigure[Al$_{17}$-O$_{3}$-Al$_{17}$\label{fig:O3vib}]{
\includegraphics[width=0.3\linewidth]{VibrationsO3}
}
\caption[Al$_{17}$-O$_{x}$-Al$_{17}$: Vibrational mode energy]
{\label{fig:Vibrationen}
Al$_{17}$-O$_{x}$-Al$_{17}$: Energy of the 
vibrational modes of the oxygen molecule connected to Al pyramids as a 
function of the distance $d$ between the Al tip atoms.   
The arrows and plus/minus signs denote the directions of relative displacements
of the atoms in a given mode:
$\rightarrow$ and $\leftarrow$ denote motion along the axis of the contact
while $+$,$-$ and $\uparrow$,$\downarrow$ denote motion 
in the two directions perpendicular to the axis.}
\end{minipage}
\end{center}
\end{figure*}

We simulate the opening of a contact for the three most basic forms of
oxygen (O, O$_{2}$, and O$_{3}$). Fig.~\ref{fig:Struktur} shows a
selection of the relaxed geometries obtained, the starting
configuration, one halfway, the one just before, and the one directly
after the breaking.  In the case of a single O atom, the O sits at
first on the side of the Al dimer contact. Upon stretching this atom
moves onto the contact axis and stays symmetrically bonded to both
pyramids until the contact breaks [Fig.~\ref{fig:Ostruct}].  When the
contact is broken, the oxygen atom binds to one electrode with a
distance of approximately 1.7 \AA\ on top of the pyramid and stays
bonded like this.  However, this binding position is only metastable.
If one distorts the position of the oxygen atom slightly, e.g. places
it at some distance from the contact axis and optimizes the contact
geometry again, the atom goes to one side of the pyramid in the same
way as for O$_{2}$ and O$_{3}$ described below. This position with the
oxygen bonded to one side of the pyramid is energetically favored by
approximately 2.5 eV.  In the case of O$_{2}$, the molecular axis of
the oxygen molecule is in the beginning almost perpendicular to the
contact axis and rotates afterwards into this axis
[Fig.~\ref{fig:O2struct}].  For O$_{3}$, the molecule is twisted
around the contact axis. This twist contracts towards the contact axis
when the contact is stretched [Fig.~\ref{fig:O3struct}].  In the cases
of Al-O$_{2}$/O$_{3}$-Al, the O atoms move always to the sides of the
pyramids when the contact is broken.  The evolution of the distances
between the atoms while opening the contact is shown in
Tab.~\ref{tab:distances}.

As mentioned above, in order to compare with results from
inelastic transport experiments, we also calculate the vibrational
modes of the oxygen molecules in the contact.  Since the atomic mass
of Al is almost twice as large as that of O, it is a reasonable first
approximation to neglect the motion of the Al atoms, which is what our model
assumes.  The energies of the vibrational modes for the different
structures are shown in Fig.~\ref{fig:Vibrationen}. For these
structures, the modes can be classified roughly as being longitudinal
or transverse, according to the motions of the atoms relative to the
axis of the contact.  The mode energies are in the range of 50 meV to
140 meV for Al-O-Al. The two transverse modes are more or less
constant around 50 meV, while the longitudinal mode increases from 120
meV to a maximum of 140 meV at a distance of 3.3 \AA\ and drops
afterwards almost linearly. For Al-O$_{2}$-Al, the mode energies are
in the range of 40 meV to 150 meV. The two lowest modes are transverse
and constant in energy. The highest mode is longitudinal and increases
from 135 meV to 150 meV at 4.5 \AA\ and drops afterwards to 120 meV.
The other three modes lie in the range of 70 meV to 110 meV. For
Al-O$_{3}$-Al, the lowest (transverse) modes lie constant at an energy
of approximately 25 meV. The highest (longitudinal) mode increases
first to its maximum of 160 meV at 5.4 \AA\ and decreases afterwards.
Another mode lies close to this highest one. All other modes have
energies between 50 meV and 120 meV. 

The behavior of the vibrational energies can be understood by
re-examining the atomic distances in the geometries
(Tab.~\ref{tab:distances}). Despite the fact that the atomic contacts
are being stretched, it happens for example that in the beginning some
atoms move closer together.  Therefore certain bonds strengthen due to
this bond shortening, and corresponding modes may show an initial
increase in their vibrational frequencies.

\section{Transport properties and electronic 
structure \label{chap:transport}}

\begin{figure}
\begin{center}
\begin{minipage}{0.96\linewidth}
\includegraphics[width=\linewidth]{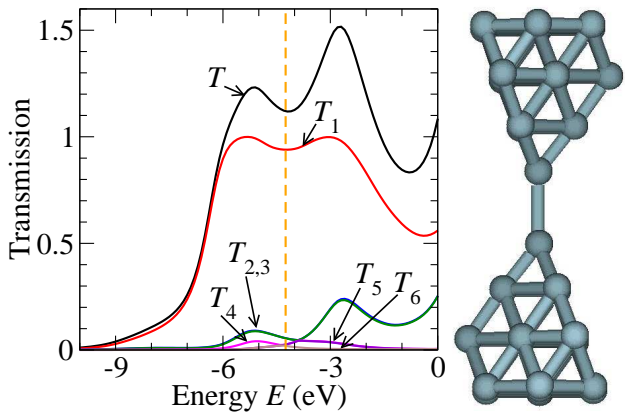}
\caption[Al-Al: Transmission] {\label{fig:TransmissionAl} (Color
  online) Al-Al: Transmission as a function of energy. Both the total
  transmission ($T$) as well as its channel contributions ($T_n$) are
  shown. The vertical dashed line indicates the Fermi energy of bulk
  Al. The corresponding contact geometry with a distance between the
  Al tip atoms of 3.4~\AA\ is depicted to the right.}
\end{minipage}
\end{center}
\end{figure}

\begin{figure*}
\begin{center}
\begin{minipage}{0.96\linewidth}
\includegraphics[width=0.7\linewidth]{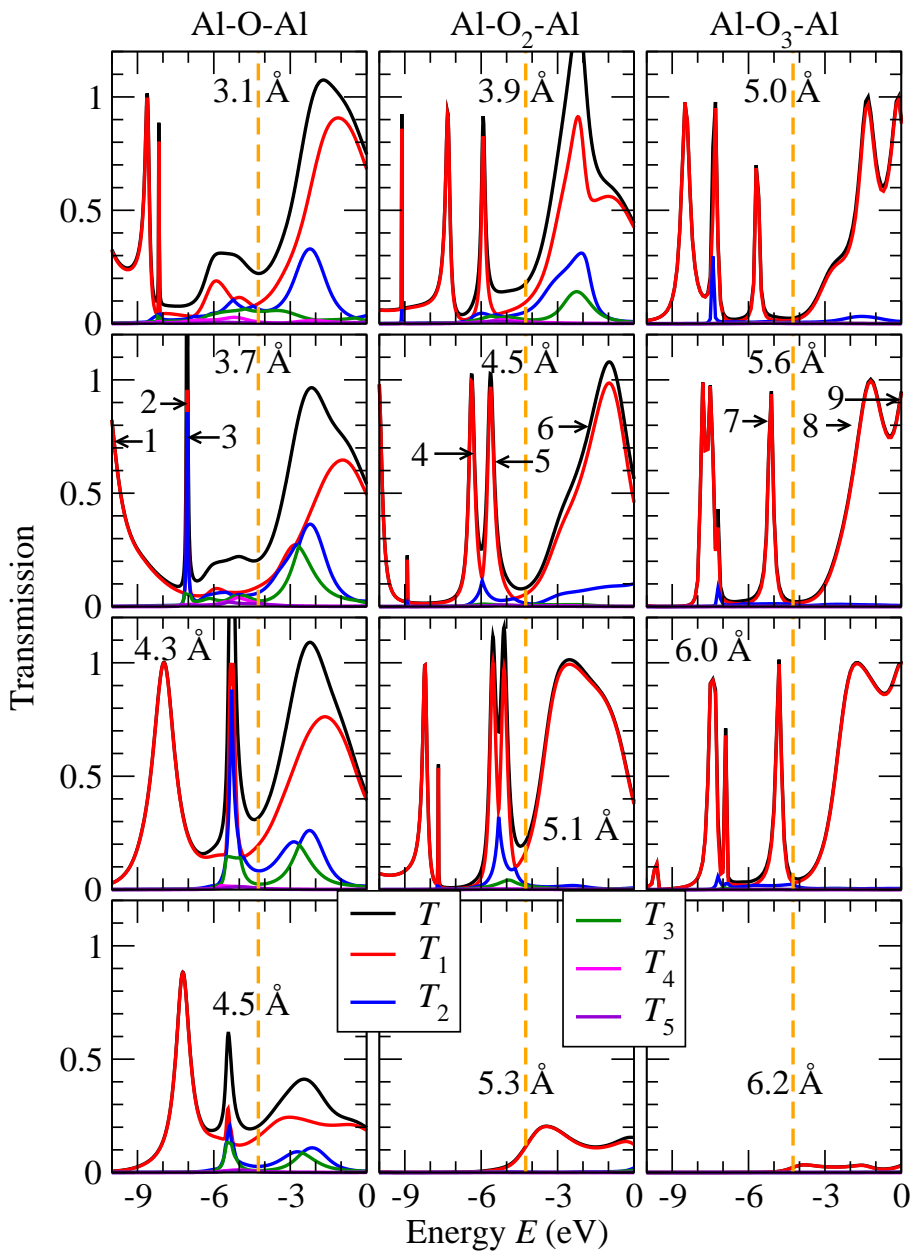}
\caption[Al-O$_\textrm{x}$-Al: Transmission]
{\label{fig:TransmissionOx}
(Color online)
Al-O$_\textrm{x}$-Al: 
Transmission as a function of energy for the structures shown in 
Fig.~\ref{fig:Struktur}. The tip-to-tip distance increases from the top to the 
bottom and is given in the panels. 
The vertical dashed lines indicate the Fermi energy.}
\end{minipage}
\end{center}
\end{figure*}

Before we determine the transport properties for the different forms
of oxygen between aluminum electrodes we will show the results for a
pure Al atomic contact. The transmission as a function of energy and
the corresponding contact geometry are plotted in
Fig.~\ref{fig:TransmissionAl}.  The conductance of this contact is
close to $1 G_0$ and is carried by 3 channels, as
expected.\cite{Scheer1997,Cuevas1998,Cron2001,Pauly2006} If oxygen
sits between the two electrodes, the transmission changes
significantly. The resulting $T(E)$ curves for the contact geometries of
Fig.~\ref{fig:Struktur} are shown in Fig.~\ref{fig:TransmissionOx}.
We note that whenever we speak about numbers of channels in this work,
we mean channels that contribute at least 5\% to the total
conductance. This criterion approximately agrees with what can be
resolved experimentally using superconducting
electrodes.\cite{Scheer1997}

\begin{figure*}
\begin{center}
\begin{minipage}{0.96\linewidth}
\includegraphics[width=0.9\linewidth]{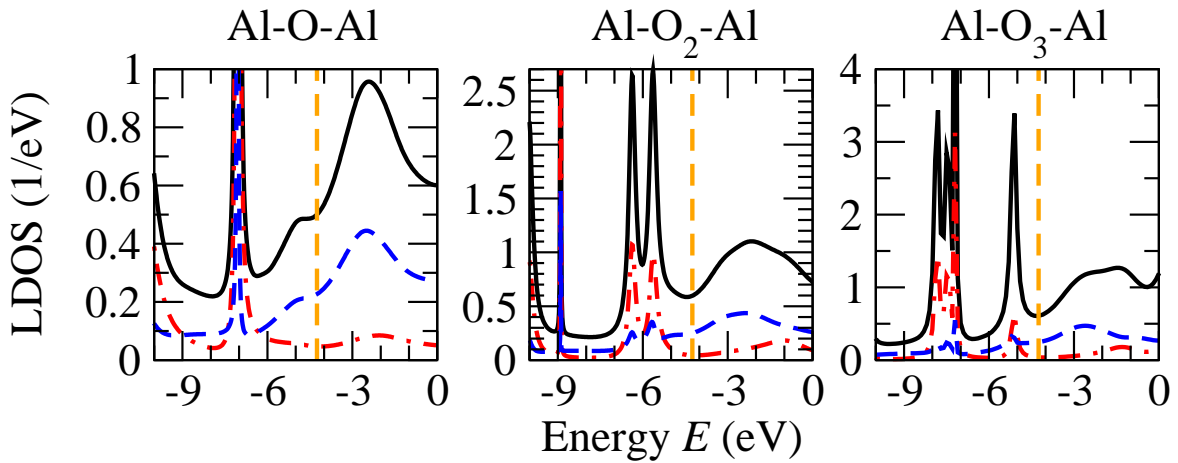}
\caption[Al-O$_\textrm{x}$-Al: LDOS.]
{\label{fig:LDOSOx}
(Color online)
Al-O$_\textrm{x}$-Al: 
LDOS corresponding to the second row in Fig.~\ref{fig:TransmissionOx}. 
The solid (black) curve is for the region including the Al tip atoms and 
the O atoms, the dash-dotted (red) one is for one of the oxygen atoms, 
and the dashed (blue) curve is for one of the
Al tip atoms. The vertical dashed (orange) line
indicates the Fermi energy.}
\end{minipage}
\end{center}
\end{figure*}

\begin{figure*}
\begin{center}
\begin{minipage}{0.9\linewidth}
\includegraphics[width=\linewidth]{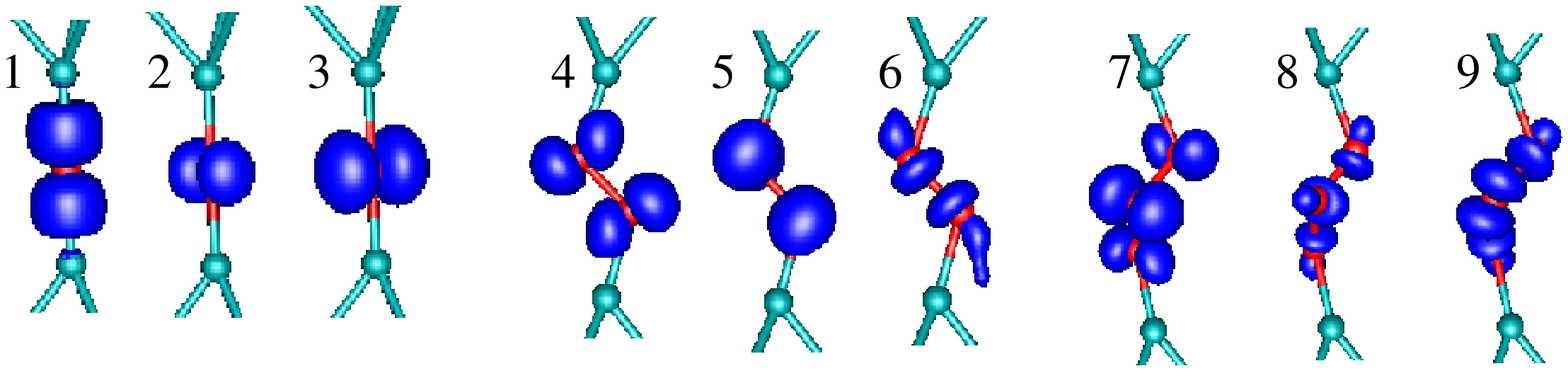}
\caption[Al$_{17}$-O$_x$-Al$_{17}$: Orbitals.]
{\label{fig:Orbitals}
(Color online)
Al$_{17}$-O$_x$-Al$_{17}$: 
Orbitals of the contact geometries of Fig.~\ref{fig:Struktur}. 
The orbitals 1--3 belong to Al$_{17}$-O-Al$_{17}$ with a tip-to-tip
distance of $d=3.7$ \AA , the orbitals 4--6 to Al$_{17}$-O$_2$-Al$_{17}$ 
with $d=4.5$ \AA, and the orbitals 7--9 to Al$_{17}$-O$_3$-Al$_{17}$
with $d=5.6$ \AA. The energies of the orbitals are 
-10.537 eV (1), -7.238 eV (2), -7.176 eV (3), 
-6.274 eV (4), -5.606 eV (5), -2.133 eV (6), -5.197 eV (7), -2.175 eV (8), 
and +0.153 eV (9), where the number of the orbital is given in brackets.}
\end{minipage}
\end{center}
\end{figure*}

The total transmissions $T(E)=\sum_nT_n(E)$ and the transmission
eigenvalues $T_{n}(E)$ of the individual channels of Al-O-Al
[Fig.~\ref{fig:Ostruct}] are plotted as a function of energy $E$ in
the left column of Fig.~\ref{fig:TransmissionOx}. The peaks labeled
1--3 below the Fermi energy are due to oxygen. This can be seen from
the local density of states (LDOS) shown for a tip-to-tip distance 
of 3.7 \AA\ in Fig.~\ref{fig:LDOSOx}  (Ref.~\onlinecite{noteLDOS})
as well as from the orbitals at the peaks shown in
Fig.~\ref{fig:Orbitals}.  Peak number 1 corresponds to the $p_z$
orbital of O ($z$ is the transport direction). Peaks 2 and 3 become
almost degenerate as soon as the O is on the contact axis. These peaks
are due to the $p_x$ and $p_y$ orbitals of O. Peak 1 is broader and
shifted to lower energies than peaks 2 and 3 due to the stronger
binding of the $p_z$ orbital to the Al states.  The peaks above the
Fermi energy are due to aluminum.  The number of channels contributing
to the conductance is at least three. When the contact is opened,
peaks 1--3 move, up to a tip-to-tip distance of 3.3~\AA, away from 
and afterwards towards the Fermi energy. This behavior is caused 
by the length of the Al-O bond.

For Al-O$_{2}$-Al [Fig.~\ref{fig:O2struct}], the total transmission
and the channel transmissions are shown in the central column of
Fig.~\ref{fig:TransmissionOx}.  The peaks in $T(E)$ are caused by the
oxygen as can be seen from the LDOS in Fig.~\ref{fig:LDOSOx} for a
tip-to-tip distance of 4.5~\AA. Peaks number 4 and 5 originate from
the anti-bonding $\pi$ orbitals and peak 6 from the anti-bonding
$\sigma$ orbital (Fig.~\ref{fig:Orbitals}). The number of channels at
the Fermi energy is two for the intermediate tip-to-tip distances
between 4.6~\AA\ and 4.9~\AA, but otherwise three.  Peaks 4 and 5
shift upwards in energy without crossing the Fermi energy while the
contact is stretched. Peak 6 moves first away from and then, after a
tip-to-tip distance of 4.5~\AA, towards the Fermi energy. The movement
of this peak can be understood in terms of the behavior of the
distance between the O atoms, while the length of the Al-O bond is
mainly responsible for the displacement of peaks 4 and 5.

For Al-O$_{3}$-Al [Fig.~\ref{fig:O3struct}], the transmissions are
plotted in the right column of Fig.~\ref{fig:TransmissionOx}.  The
peaks shown have their origin in the ozone (Fig.~\ref{fig:LDOSOx}
and~\ref{fig:Orbitals}). Peaks 8 and 9 move away from the Fermi energy
with stretching up to a tip-to-tip distance of 5.4~\AA\ and approach
afterwards the Fermi energy. Peak 7 shifts continuously towards the
Fermi energy.  The movement of peaks 8 and 9 again reflects the
distances between the oxygen atoms, and the shift of peak 7 changes in
the Al-O bond length. The current is carried, up to a tip-to-tip
distance of 5.4~\AA, by three channels and afterwards by two channels.

\begin{figure}
\begin{center}
\begin{minipage}{0.9\linewidth}
\includegraphics[width=\linewidth]{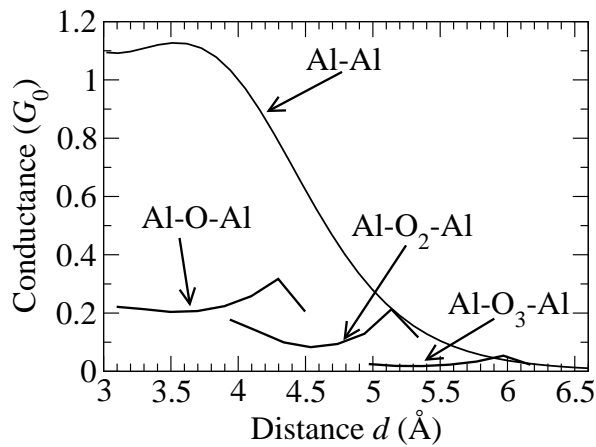}
\caption[Al-O$_{x}$-Al: Conductance.]
{\label{fig:Conductance}
Al-O$_{x}$-Al: 
Conductance as a function of the distance between the Al tip atoms.}
\end{minipage}
\end{center}
\end{figure}

In Fig.~\ref{fig:Conductance} the conductance is plotted as a function
of the tip-to-tip distance for pure Al and for the three different
oxygen molecules in Al contacts. For pure Al contacts we obtain a
conductance close to $1 G_0$ with a positive slope shortly before
rupture of the contact, in agreement with previous investigations of
Al atomic contacts.\cite{Krans1993,Scheer1997,Jelinek2005,Pauly2006}
In the presence of oxygen, the conductances show the same qualitative
features for the three molecules (O, O$_2$, and O$_3$).  The
conductance first decreases with stretching up to a certain point.
After this point it increases rapidly until the contact breaks and the
conductance drops. This trend can be understood by re-examining the
evolution of the energy dependence of the transmission upon stretching
(Fig.~\ref{fig:TransmissionOx}). Thus for instance, the increase of
the conductance originates from the fact that the highest occupied
molecular orbitals move close to the Fermi energy.  Quantitatively the
conductance depends on the number of oxygen atoms in the molecule. The
more oxygen atoms, the lower the conductance. In all cases, the
transport is influenced by a charge transfer from the Al to the oxygen
molecule (around $-0.6e$ for all the three molecules, as determined by
a Mulliken population analysis) leading to an occupation of
molecular orbitals which would be empty in an isolated oxygen
molecule. Therefore, current is carried by formerly unoccupied $p$
orbitals.

We also investigate the conductance in the tunneling regime.  With
this term we refer to the region of large tip-to-tip distances $d$,
where the conductance exhibits an exponential decay 
$G(d)/G_0\propto e^{-\beta d}$ with an inverse decay length $\beta$.
In experiments there might be any number of O
molecules covering the surfaces of the Al pyramids, but we calculate
representatively the tunneling conductance of Al-O$_2$-Al (not shown
in Fig.~\ref{fig:Conductance}), where the O atoms sit separated on
both pyramids [see the rightmost part of Fig.~\ref{fig:O2struct}]. We
compare this to the tunneling conductance of pure Al contacts (shown
in Fig.~\ref{fig:Conductance}). For pure Al, the inverse decay length 
turns out to be $\beta = 2.1$ \AA $^{-1}$ and for Al with an O atom
on both pyramids $\beta = 2.3$ \AA $^{-1}$.  Thus the decay of the
conductance is somewhat faster when oxygen is present.

Physically, the value of $\beta$ is determined by the shape of the
potential barrier between the tips, especially its apparent height
with respect to $E_F$ (Ref.\ \onlinecite{Lang1988}).  The barrier
depends sensitively on the electronic structure of the junction,
including the charge-transfer effects between the different atoms
close to the tips.  In terms of the local-basis picture, all this
information is contained in the spectral functions of the tips and the
hopping matrix elements connecting them.  The mathematical details of
this interpretation are discussed further in App.\ \ref{app.a}. In
this picture, an important role is played by the radial decay
properties of the orbitals of the tip atoms that are relevant for
transport. Roughly speaking, for O these orbitals are the 2$p$ ones,
while for Al they are of the 3$s$ and 3$p$ type. The former orbitals
decay faster and thus it is reasonable that inverse decay length
$\beta$ is larger in the presence of oxygen.  Although the
oxygen-induced change in $\beta$ in this particular example is not
very large, we suggest that, in general, changes in the decay rate of
the tunneling conductance may provide another possibility to check
experimentally whether gas molecules are present in the atomic contact
or not.

\section{Discussion and conclusions 
\label{chap:discussion}}

\begin{figure}
\begin{center}
\begin{minipage}{0.9\linewidth}
\includegraphics[width=0.9\linewidth]{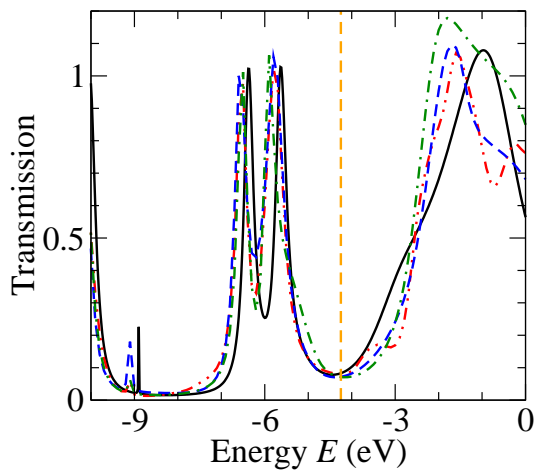}
\caption[Al-O$_{2}$-Al: compare size and cut]
{\label{fig:comparesizeandcut}
(Color online)
Robustness of the transmission for an Al-O$_{2}$-Al contact at a tip-to-tip
distance of $4.5$~\AA\ [Fig.~\ref{fig:Struktur}(b)]. 
Shown are $T(E)$ curves for an increased number of Al atoms in the pyramids
as well as for different partitionings of the contact into $L$, $C$, and $R$
regions. The solid black line is the transmission of 
7-6-$|$3-1-O$_{2}$-1-3$|$-6-7 
(Fig.~\ref{fig:TransmissionOx}). The numbers stand for the Al 
atoms in each layer of the pyramid and ``$|$'' indicates how the system
is divided into the three different regions. 
The dash-dot-dotted line corresponds to the system
6-12-$|$10-6-3-1-O$_{2}$-1-3-6-10$|$-12-6,
the dashed line to  6-12-10-$|$6-3-1-O$_{2}$-1-3-6$|$-10-12-6, and the
dash-dotted line to 6-12-10-6-$|$3-1-O$_{2}$-1-3$|$-6-10-12-6.
The vertical dashed line indicates the Fermi energy.  
}
\end{minipage}
\end{center}
\end{figure}

Let us point out that we have checked that the results presented in
this work are robust with respect to the size of the contacts. In
particular, the charge transport characteristics do not change
significantly when the size of the pyramids in the contact geometries
is increased, nor do they depend sensitively on the partitioning into
the $L$, $C$, and $R$ regions. We illustrate this fact in
Fig.~\ref{fig:comparesizeandcut}, where we compare the transmission
functions $T(E)$ of the Al-O$_{2}$-Al contact of
Fig.~\ref{fig:O2struct} at a tip-to-tip distance of 4.5 \AA.  We vary
the numbers of layers in the Al pyramids and divide the system
differently into the three regions.  Clearly, the essential features
of the results remain unaffected by these modifications. Such a
robustness is observed as long as the assumption about vanishing
matrix elements between the $L$ and $R$ regions is fulfilled, and
there are enough layers (at least two in the case of Al) left in the
$L$ and $R$ parts of the pyramids for coupling the $C$ part to the
surfaces.

\begin{figure}
\begin{center}
\begin{minipage}{0.9\linewidth}
\includegraphics[width=\linewidth]{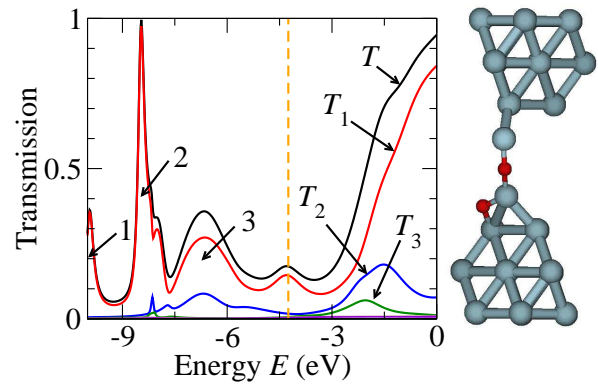}
\caption[Al-O$_{2}$-Al: Dissociated structure.]
{\label{fig:disso}
(Color online)
Al-O$_{2}$-Al: Example of a possible geometrical configuration when 
the Al tip atoms are included in the relaxation process 
in addition to the molecule, and the corresponding transmission as a
function of energy. The vertical dashed line indicates the Fermi
energy.}
\end{minipage}
\end{center}
\end{figure}

In order to investigate structures where the effect of O molecules on
the transport properties is at its largest, we have placed the oxygen
molecules initially between the pyramids and kept the Al electrodes
fixed in the geometry-optimization process.  On the other hand, we
know that relaxing parts of the pyramids together with the molecule
can lead to geometrical structures differing considerably from the
ideal contacts shown in Fig.\ \ref{fig:Struktur}. An example of such a
geometry is depicted in Fig.~\ref{fig:disso}. Moreover, the formed
structures depend on the initial placements of the atoms close to the
tips and on the number of oxygen molecules present.  Different
realizations of atomic configurations can also be observed in
experiments, where various kinds of $I$-$V$ characteristics have been
measured.\cite{Isambert2007} However, as the transmission function in
Fig.~\ref{fig:disso} shows, even more complex geometries can exhibit
results whose main features resemble those of the simplified
structures discussed above.  The transmission function shown in this
figure should be compared with the ones for the Al-O-Al contact in
Fig.\ \ref{fig:TransmissionOx}.  In particular, similarly to those
results, there are three peaks (1-3) below the Fermi energy and the
conductance is around $0.2 G_0$.

A similar approach to investigate oxygen in Al junctions was chosen by
Jel\'{\i}nek \emph{et al.}\ (Ref.\ \onlinecite{Jelinek2005}). 
Those authors found structures where an oxygen atom is
incorporated into the electrodes and their electrodes are connected by
an aluminum dimer in the final stage before the contact breaks.  The
conductance in their structures, therefore, is dominated by the
aluminum dimer and the influence of the oxygen atom on the conductance
is weak. We nevertheless believe that structures with oxygen molecules
bridging the gap between Al electrodes can be found in 
experiments. A situation corresponding to our simulations would most
likely arise when the oxygen is introduced into open contacts that are
subsequently closed and then reopened. The experimental results can
then be compared to our predictions.

In conclusion, we studied the charge transport properties of aluminum
atomic contacts in the presence of oxygen molecules (O, O$_{2}$, and
O$_{3}$).  We obtained stable geometrical structures for the molecules
in the junctions, and determined the molecules' vibrational modes in
the presence of the aluminum electrodes. In principle these modes
should be measurable with the methods of point contact spectroscopy.
Moreover, we analyzed the evolution of the conductance for the
Al-O/O$_{2}$/O$_{3}$-Al contacts in stretching processes, mimicking a
conductance curve measured in experiments while opening a break
junction.  These results show a typical behavior for all the different
molecules studied. First the conductance decreases, until it starts to
increase shortly before the contact breaks. The value of the
conductance was found to decrease with the number of oxygen atoms. Our
observations can be understood by the interplay between mechanical and
electronic properties.  Changes in the distances between the atoms
result in modifications of the electronic structure. The shifts in
energy positions of individual molecular orbitals and their
corresponding peaks in the transmission function can in turn explain
the behavior of the conductance in the stretching processes.  Finally,
we studied the decay lengths of the conductance in the tunneling
regime, and found the decay to be faster in the presence of oxygen
than for pure aluminum contacts. This effect may provide a way for
detecting gas molecules in metallic atomic contacts.

\begin{acknowledgments}
We would like to thank Michael H\"{a}fner, Marcelo
Goffman, and Antoine Isambert for sharing their ideas
with us. We also thank the Theoretical Chemis\-try
group of Reinhart Ahlrichs for helpful discussions
and for providing us with \textsc{Turbomole}. Financial
support by the Helmholtz Gemeinschaft within the
``Nachwuchsgruppen-Programm'' (Contract No.\ VH-NG-029),
the Landesstiftung Baden-W\"{u}rttemberg within the
``Kompetenznetz Funktionelle Nanostrukturen'', and the
DFG within the CFN is gratefully acknowledged.

\end{acknowledgments}

\appendix

\section{Conductance formula in the tunneling limit}
\label{app.a}

In the actual conductance calculations presented in this paper, we
always use the formula given in Eq.\ (\ref{e.landauer}).  For a better
understanding of the tunneling regime, it is nevertheless useful to
cast Eq.\ (\ref{e.landauer}) into another well-known form.  Let us
assume that the $C$ part of the system can be divided into two regions
1 and 2, where region 1 (2) is not coupled to the $R$ ($L$) lead through
direct hoppings or overlaps.  Furthermore, regions 1 and 2 are connected
to each other by 
$\omat{t}_{12}=\omat{H}_{12}-E\omat{S}_{12}$.  Equation
(\ref{e.landauer}) can then be written in the form
\begin{equation}\label{e.landauer2}
G=G_0\textrm{Tr}[
\omat{G}^r_{12}(\omat{\Gamma}_R)_{22}
\omat{G}^a_{21}(\omat{\Gamma}_L)_{11}],
\end{equation}
where $\omat{G}^a_{21}=(\omat{G}^r_{12})^\dagger$, and all energy-dependent 
quantities are evaluated at $E_F$. 
Let us denote by 
$\omat{g}^r_{11}$
and
$\omat{g}^r_{22}$
the Green's functions for the 
regions 1 and 2 in the absence of $\omat{t}_{12}$. 
For example
$\omat{g}^r_{11}(E)=[E\omat{S}_{11}-\omat{H}_{11}-(\omat{\Sigma}^r_L)_{11}]^{-1}$.
We also define
the unperturbed spectral density matrix
$\omat{A}_{11}= i(\omat{g}^r_{11}-\omat{g}^a_{11})$, where
$\omat{g}^a_{11}=(\omat{g}^r_{11})^\dagger$,
with a similar definition for $\omat{A}_{22}$.
We note that $\omat{H}_{11}$ and $\omat{H}_{22}$ 
are simply cut out from the full, self-consistent $\omat{H}_{CC}$
and thus still contain indirect effects of $\omat{t}_{12}$.

The full propagator between 1 and 2 is given by the 
Dyson equation\cite{Williams1982}
$\omat{G}^r_{12}=\omat{g}^r_{11}\omat{t}_{12}\omat{G}^r_{22}$
$=\omat{g}^r_{11}\omat{T}_{12}\omat{g}^r_{22}$.
In the last stage, we defined the $t$ matrix $\omat{T}_{12}$
due to the perturbation $\omat{t}_{12}$ (Ref.\ \onlinecite{Galperin1999}).
Inserting these into Eq.\ (\ref{e.landauer2}) and using
the identities 
$\omat{g}^r_{11}(\omat{\Gamma}_{L})_{11}\omat{g}^a_{11}=\omat{A}_{11}$ and
$\omat{g}^r_{22}(\omat{\Gamma}_{R})_{22}\omat{g}^a_{22}=\omat{A}_{22}$, 
we find the following formula\cite{Todorov1993}
\begin{equation}\label{e.tunnel2}
G=G_0\textrm{Tr}[
\omat{T}_{12}\omat{A}_{22}
\omat{T}_{21}\omat{A}_{11}].
\end{equation}
Far in the tunneling regime, the quantities $\omat{t}_{12}$ are small,
such that the lowest-order approximation $\omat{T}_{12}=\omat{t}_{12}$
should be valid.  Furthermore, to lowest order, the electronic
structures of the tips (as represented by $\omat{H}_{11}$ and
$\omat{H}_{22}$, and hence $\omat{A}_{11}$ and $\omat{A}_{22}$) 
also become independent of the tip-to-tip distance. 

The decay rate of the conductance with increasing distance between the
tips ($\beta$) is now seen to be determined by two factors. The first
is due to the decay of the couplings $\omat{t}_{12}$.  If one uses the
Wolfsberg-Helmholz (or ``extended H\"uckel'') approximation, then
$\omat{H}_{12}$ and hence $\omat{t}_{12}$ are related in a simple way
to the overlap $\omat{S}_{12}$ (Ref.\ \onlinecite{Wolfsberg1952}).
This emphasizes that the major effect to the decay of $\omat{t}_{12}$
is simply due to the associated local basis functions, which decay
with varying rates.  The second factor is due to the spectral
functions $\omat{A}_{11}$ and $\omat{A}_{22}$.  They weight the
various components of $\omat{t}_{12}$ differently, and thus finally
determine the decay rate $\beta$.  The physical quantity controlling
this rate is still the effective tunneling barrier, in particular its
height with respect to $E_F$, because the latter determines the
barrier-penetration lengths of the electronic eigenstates of the tips.
This information is contained in the matrices $\omat{A}_{11}$ and
$\omat{A}_{22}$ in the form of the weights of the different local
basis functions of the tip atoms.

\end{document}